\documentclass[review]{elsarticle}

\usepackage{amsmath}
\usepackage{amssymb}
\usepackage{graphicx}
\usepackage{enumerate}
\usepackage{algorithm}
\usepackage{amsthm}
\usepackage[noend]{algpseudocode}

\usepackage[margin=3.5cm]{geometry}

\makeatletter
\def\BState{\State\hskip-\ALG@thistlm}
\makeatother

\journal{Arxiv}

\biboptions{authoryear}

\begin{document}

\begin{frontmatter}

\title{Classification of Computer Models with Labelled Outputs}

\author[]{Louise Kimpton\corref{cor1}}
\ead{lmk212@exeter.ac.uk}

\author{Peter Challenor}
\ead{P.G.Challenor@exeter.ac.uk}

\author{Daniel B Williamson}
\ead{D.Williamson@exeter.ac.uk}

\cortext[cor1]{Corresponding author}

\address{Department of Mathematics, College of Engineering, Mathematics and Physical Sciences, University of Exeter}

\begin{abstract}
Classification is a vital tool that is important for modelling many complex numerical models. A model or system may be such that, for certain areas of input space, the output either does not exist, or is not in a quantifiable form. Here, we present a new method for classification where the model outputs are given distinct classifying labels, which we model using a latent Gaussian process (GP). The latent variable is estimated using MCMC sampling, a unique likelihood and distinct prior specifications. Our classifier is then verified by calculating a misclassification rate across the input space.  

Comparisons are made with other existing classification methods including logistic regression, which models the probability of being classified into one of two regions. To make classification predictions we draw from an independent Bernoulli distribution, meaning that distance correlation is lost from the independent draws and so can result in many misclassifications. By modelling the labels using a latent GP, this problem does not occur in our method. We apply our novel method to a range of examples including a motivating example which models the hormones associated with the reproductive system in mammals, where the two labelled outputs are high and low rates of reproduction. 
\end{abstract}


\end{frontmatter}


\section{Introduction}

It is common to use complex numerical models to represent real life physical systems. By using simulators or models, we can reproduce data, make predictions and generally get a better understanding of these complex systems \citep{Sacks1989}. For simulators that can be time consuming or expensive to run, there are statistical approaches known as emulators which act as a `black box' model to represent statistically the relationships between the simulator inputs and outputs \citep{Kennedya}. Emulators thus provide a deeper understanding of the complex interactions involved in the physical systems, as well as defining any uncertainty.

In some numerical models, the model output can be labelled as belonging to two or more classes and a classification method is needed to split the input space up into regions according to these labelled outputs. The model outputs can take any form and be either qualitative or quantitative. Possible outputs could be $\{\text{high},\text{low}\}$, $\{\text{red},\text{green}\}$ or just $\{0,1\}$ for our two separate regions. For example, we may have computer code for a complex model that fails to run for certain input values \citep{Edwards2011}. In such an example, our data would correspond to separate binary outcomes of `runs' and `fails to run', and we could allocate a separate class label for each of these outcomes. Thus, by using a separate labelling variable (we shall refer to this as class labelling), we can ignore any output function (if it exists) to make sure we can properly address these types of problems.

Whilst classical classification exists for sorting data into specified regions, it often neglects any information regarding distance in the input space. Continuing our fail/not fail example from above, if we knew one input value where a model is certain to `fail' to compute, it is sensible to assume that other similar input values are also likely to fail. This spatial relationship between neighbouring points is valuable and should be incorporated into our classification. The aim of this paper is to develop a new method of uncertainty quantification (UQ) classification for computer models that have two distinct labelled regions, and where an output function is not necessarily quantitative across the whole input space. Class labelling will be a vital aspect of our proposed method since (as we described) some applications do not have quantitative simulator or model function outputs.

To retain a distance measure between inputs in our method when we classify, we define that our input space lives in a vector space. This is not the case for current methods such as logistic regression \citep{Diggle1998, Chang2015} and machine learning classification \citep{Seeger2004, Nickisch2008, Chan2013}. 

If we were to use logistic regression, we would produce a posterior distribution for the predictive class membership of being in one of the two regions, which can include distance information. However, when we sample from the distribution, or use it to make predictions, we draw from an independent Bernoulli distribution where the 0/1 outputs correspond to either of the two regions. Drawing marginally in this way, instead of from a joint distribution means that any correlation between inputs is lost, causing far more misclassifications to happen. We discuss this further in Section \ref{comp}.

Take a simple example, similar to that mentioned by \cite{Chang2015} concerned with classifying areas of ice sheet and ocean. We assume just one spatial input along a line where we know whether there is any ice sheet present at four initial points. The first two points are known to definitely be ice (region 1) and the following two points are known to be ocean (region 2). Logically, there is a much higher chance of finding ice sheet close to where existing points are already known to be ice (region 1), rather than ocean (region two). If we drew points independently close to these known points, then there is still a chance that we may result in a misclassification; it is thus important to include some correlation over distance in our model.

Suppose we restrict this example with the advanced knowledge that there is only one boundary between ice and ocean in our input space. The change in label can happen anywhere between the two central points and we assume a hard boundary (i.e. in any realisation any point is either ice or ocean). The input space between each pair of points in the same regions, however, must be classified with the same label as the surrounding points. As we get closer to this boundary, the probability of being classified into the first region becomes close to 50\% since we are uncertain of where the exact boundary lies. Hence, the draws from the Bernoulli distribution become equally likely to fall on a 0 or a 1, and so there will be a section (close to the boundary) where the classification may appear random.  Our draws are then unrealistic, because we know a hard boundary exists.

\cite{Ranjan2008} proposed an alternative method to classification and logistic regression by modelling the boundary between the two separate output regions specifically as a contour. They attempt to estimate the contour defining the outputs of a complex computer code based on an improvement function. Although their method appears to be an improvement in producing uncertainty, it requires an underlying smoothness assumption, as the whole output space is modelled by a single Gaussian process. As such, this method is unsuitable for our fail/not fail example. It is also likely to become increasingly complex in higher dimensions.

A process known as history matching is used in a method developed by \cite{Caiado2015}. History matching is an iterative process designed to reduce the input space of the simulator such that input values that are not likely to result in the observed data are discarded \citep{Andrianakis2015, Vernon2010, Salter2018}. In particular, \cite{Caiado2015} use it to sort data into the separate output regions by discarding regions which are unlikely based on an implausibility criteria. Although this has no smoothness assumption, it may develop added complexities in higher dimensions.

We present here a new method for this problem that encapsulates the ideas of general classification and UQ, extending it to a wider range of applications, and includes distance correlation between our model inputs. We have included ideas from \cite{Nickisch2008} and \cite{Chang2015} to model the regions using class labels in latent space with a Gaussian process. A motivational example has been supplied by \cite{Voliotis2018}, who model the reproduction system in mammals. The two dimensional model consists of a set of coupled ordinary differential equations where the inputs are certain hormones linked to the causes of high and low rates of reproduction. This creates a system with two distinct solutions (high and low rates). Being able to accurately model their system, and locate the areas of low and high firing rates means that not only can we aid predictions on the reproduction rate, but we can also have a better understanding of the specific input parameters that are associated with the different rates of reproduction.

In Section \ref{meth}, we give an outline of our method including a brief overview of Gaussian processes. In Section \ref{1de}, we apply our method to a simple one dimensional example. Section \ref{miss} then discusses our approach to model validation. In Section \ref{prior}, we explain some of the prior choices that are vital to our method. Section \ref{2de} extends our method to a two dimensional example, followed by some comparisons with existing methods in Section \ref{comp} and a more complicated example in Section \ref{Sant}. The motivational example is outlined in Section \ref{app}. Finally in Section \ref{con}, we conclude with a discussion and overview.

\section{Methodology} \label{meth}
Let $\mathcal{X}$ be a normed vector space with norm $\| . \|$. Let $\textbf{x} = \textbf{x}_{1}, \ldots, \textbf{x}_{n} \in D$ be inputs to a model in $p$ dimensions that lie within $\mathcal{X}$. The input space, $D$, is partioned into 2 regions, $R_{1}$ and $R_{2}$, such that $R_{1} \cup R_{2} = D$ and $R_{1} \cap R_{2} = \emptyset$.

The function, $f(.)$, that maps the inputs, $\textbf{x} \in D$, to the outputs of the model, $f(\textbf{x})$, may lie in real (or complex) space, but may also be qualitative (e.g. fail/not fail) or simply take values $f(\textbf{x}) \in \{0,1\}$. For generality of our method, we therefore define the function $\Lambda(.)$ which assigns a class labelling to each of the input data points, $\textbf{x}_{1}, \ldots, \textbf{x}_{n}$ as follows:
\begin{equation}
\begin{split}
\Lambda : D \longmapsto \{l_{1},l_{2}\} ; \hspace{0.2cm} \Lambda(\textbf{x}) &= l_{1} \hspace{0.5cm} \forall \hspace{0.3cm} \textbf{x} \in R_{1} \\
\Lambda(\textbf{x}) &= l_{2} \hspace{0.5cm} \forall \hspace{0.3cm} \textbf{x} \in R_{2} .
\end{split}
\end{equation}

For an example of application, recall the example where we have a computer model that only runs to completion for certain inputs, $\textbf{x}$. The simulator output here takes values $f(\textbf{x}) \in \{\text{fail}, \text{not fail}\}$. The inputs that lead to a failed run will lie in $R_{1}$ and be given label $l_{1}$, whilst all inputs that do run to completion will lie in $R_{2}$ and be given label $l_{2}$.

We model $\Lambda(\textbf{x})$ using a latent Gaussian process (GP), $\eta(\textbf{x})$, so that:
\begin{equation}
\Lambda(\textbf{x}) | \eta(\textbf{x}) = 
\begin{cases}
\hspace{0.2cm} l_{1} \hspace{0.3cm} &\forall \hspace{0.2cm} \textbf{x} \in D \hspace{0.1cm} : \hspace{0.1cm} \eta(\textbf{x}) < 0 \\
\hspace{0.2cm} l_{2} \hspace{0.3cm} &\forall \hspace{0.2cm} \textbf{x} \in D \hspace{0.1cm} : \hspace{0.1cm} \eta(\textbf{x}) \ge 0 ,
\end{cases}
\end{equation}
$\text{and} \hspace{0.3cm} \eta(\textbf{x}) \sim GP(m(\textbf{x}), v(\textbf{x},\textbf{x}'))$

A Gaussian process is a generalisation of the normal distribution to infinite dimensions that defines a distribution over functions \citep{Kennedya}. Any finite collection of random variables from a Gaussian process has a multivariate Normal distribution. They are fully defined by their mean function, $m(.)$, and covariance function, $v(.,.)$, where:
\begin{equation}
\begin{split}
m : D \longmapsto \mathbb{R} ; \hspace{0.2cm} m(\textbf{x}) &= \mathbb{E}[\eta(\textbf{x})] \\
v : D \times D \longmapsto \mathbb{R} ; \hspace{0.2cm} v(\textbf{x},\textbf{x}') &= \text{Cov}[\eta(\textbf{x}),\eta(\textbf{x}')] .
\end{split}
\end{equation}

The mean function allows us to input our prior belief about the form of $\eta$. Here we will restrict ourselves to Gaussian processes with linear prior mean functions, specified in the form: $\mathbb{E}[\eta(\textbf{x})|\beta] = \textbf{h}(\textbf{x})^{T}\beta$, where $\textbf{h}(\textbf{x})$ is a vector of basis functions of $\textbf{x}$ and $\beta$ is a vector comprising of unknown coefficients. We assume that the latent Gaussian process is stationary such that $\text{Cov}[\eta(\textbf{x}),\eta(\textbf{x}')]$ is a function of distance $\| \textbf{x} - \textbf{x}' \|$, where we define $\| \textbf{x} \| = \langle \textbf{x}, \textbf{x}' \rangle$, with $\langle \textbf{x}, \textbf{x}' \rangle$ being an inner product in space. We write the covariance function as $\sigma^{2}c(\textbf{x},\textbf{x}')$, where $\sigma^{2}$ is the process variance and $c$ is a known correlation function of $\| \textbf{x} - \textbf{x}' \|$. A common choice of correlation function is the squared exponential; $c(\textbf{x}_{i},\textbf{x}_{j}) = \text{exp} \left\{ - \frac{\| \textbf{x} - \textbf{x}' \| ^{2}}{\delta} \right\}$, where $\delta$ is the correlation length parameter, which controls the smoothness of the process and how much it can be perturbed as the inputs are varied. The covariance function ensures that the classification is now dependent on the distance between the inputs, $\textbf{x}$.

To update our prior Gaussian process, with data $\Lambda(\textbf{x}) = \Lambda(x_{1}), ..., \Lambda(x_{n})$, we require the likelihood, $P(\Lambda(\textbf{x}) | \eta(\textbf{x}))$, where:
\begin{equation} \label{E1}
\begin{split}
P(\Lambda(\textbf{x}) | \eta(\textbf{x})) & = P \left( \eta(x_{1})<0 , \eta(x_{2})<0 , \ldots , \eta(x_{j})<0 ,  \eta(x_{j+1})>0 , \ldots , \eta(x_{n})>0 \right) \\
& = \int_{-\infty}^{0} \dots \int_{-\infty}^{0} \int_{0}^{\infty} \dots \int_{0}^{\infty} \phi ( \eta(x_{1}), \eta(x_{2}), \dots, \eta(x_{j}), \eta(x_{j+1}), \dots, \\
& \hspace{1cm} \eta(x_{n}) )  \hspace{0.2cm} dx_{1} dx_{2} \dots dx_{n} .
\end{split}
\end{equation}
Without loss of generality, we have labelled the first $j$ points as having label $l_{1}$ and the remaining points as having label $l_{2}$. The main difference between sampling from an ordinary Gaussian process to the latent one here is the use of a joint cumulative distribution function instead of the density in the likelihood function. It is also crucial to note that this likelihood is a joint distribution (rather than a product of marginals) as our data are correlated. The form of this likelihood is analogous to the Gaussian process fitting seen in \cite{Gosling2005} and \cite{Gosling2007}. We sample from the posterior, $P(\eta(\textbf{x}) | \Lambda(\textbf{x}), \beta, \sigma^{2}, \delta)$, using an MCMC algorithm, which we will describe below. Note that by sampling from a joint distribution, we are able to classify whole sets of points in the input space simultaneously, as well as individual points. 

In order to classify a set of new points, $\textbf{x}^{*} = x_{1}^{*}, .., x_{m}^{*}$, we require joint samples from the posterior predictive distribution:
\begin{equation}
P(\eta(\textbf{x}^{*}) | \Lambda(\textbf{x})) = \int \int P(\eta(\textbf{x}^{*}) | \eta(\textbf{x}), \theta) P (\eta(\textbf{x}) | \theta, \Lambda(\textbf{x})) P(\theta | \Lambda(\textbf{x})) \hspace{0.2cm} d\eta(\textbf{x}) d\theta ,
\end{equation}
where, $\theta = (\beta, \sigma^{2}, \delta)$.

We can obtain a set of samples from $P(\theta | \Lambda(\textbf{x}))$ using Metropolis Hastings \citep{Chib2017, Gelman2013, Brooks2012}, and given a sample from $P(\eta(\textbf{x}) | \theta, \Lambda(\textbf{x}))$, we can easily sample from:
\begin{equation}
\begin{split}
\eta(\textbf{x}^{*}) | \eta(\textbf{x}),\theta \hspace{0.2cm} &\sim \hspace{0.2cm} \mathcal{MVN} \left( m^{*}(\textbf{x}^{*}), v^{*}(\textbf{x}^{*}, \textbf{x}^{*}) \right) , \\
m^{*}(\textbf{x}^{*}) &= m(\textbf{x}^{*}) + v(\textbf{x}^{*},\textbf{x}) v(\textbf{x},\textbf{x})^{-1} \left( \eta(\textbf{x}) - m(\textbf{x}) \right) , \\
c^{*}(\textbf{x}^{*},\textbf{x}^{*}) &= c(\textbf{x}^{*},\textbf{x}^{*}) - c(\textbf{x}^{*},\textbf{x}) c(\textbf{x},\textbf{x})^{-1} c(\textbf{x}, \textbf{x}^{*}) .
\end{split}
\end{equation}

Sampling from $P(\eta(\textbf{x}) | \theta, \Lambda(\textbf{x}))$, is not so straight forward. We could preform a rejection sample, similar to that seen in the ABC literature \citep{Turner2012, Wilkinson2008}, but this is extremely inefficient for even modest ensemble size $n$. The reason for this inefficiency is due to the need for every $\eta(x_{i})$ needing to agree with the respective model labels, $\Lambda(x_{i})$, simultaneously.

Our solution is to use a Gibbs sampler \citep{Brooks2012, Gilks1996}. This method makes use of the full conditional distributions by going through each variable in turn to sample from its conditional distribution whilst the remaining variables are fixed at their current values. This is possible because all variables have a Normal distribution. Algorithm \ref{GS} outlines the Gibbs sampler used to generate posterior samples from $P(\eta(\textbf{x}) | \theta, \Lambda(\textbf{x}))$. Using this method ensures that the correlation between points remains the same and computational time is saved by about a third compared to the full ABC-MCMC. 

\begin{algorithm}
\caption{Gibbs Sampling} \label{GS}
\begin{algorithmic}[1]
\State $\text{Start with initial values } \eta_{1}^{(0)}, ..., \eta_{n}^{(0)} \text{sampled from prior distribution}$
\State $\textbf{for } i = 1, 2, ... \textbf{ do}$ 
\State $\hspace{1cm} \eta_{1}^{(i)} \sim P(\eta_{1}^{(i)} | \eta_{2}^{(i-1)}, \eta_{3}^{(i-1)}, ..., \eta_{n}^{(i-1)})$
\State $\hspace{2cm} \vdots$
\State $\hspace{1cm} \eta_{j}^{(i)} \sim P(\eta_{j}^{(i)} | \eta_{1}^{(i)}, ..., \eta_{j-1}^{(i)}, \eta_{j+1}^{(i-1)}, ..., \eta_{n}^{(i-1)})$
\State $\hspace{2cm} \vdots$
\State $\hspace{1cm} \eta_{n}^{(i)} \sim P(\eta_{n}^{(i)} | \eta_{1}^{(i)}, \eta_{2}^{(i)}, ..., \eta_{n-1}^{(i)})$ 
\State $\textbf{end for}$
\end{algorithmic}
\end{algorithm}

\section{An illustrative example in 1 dimension} \label{1de}

\begin{figure}[ht]
\centering
\includegraphics[scale=0.4]{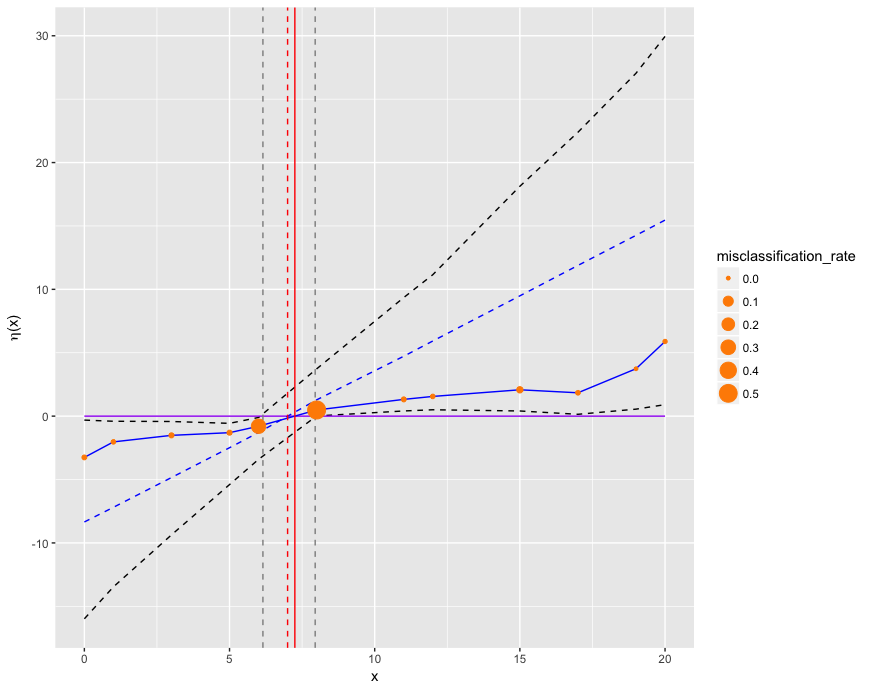}
\caption{1 dimensional example with 2 output regions. The posterior mean of the latent Gaussian process (solid blue) is shown along with the prior mean (dashed blue), true boundary (dashed red) and boundary estimate (solid red). Both have 95\% credible intervals included (black/grey dashed lines). Initial data points are shown in orange with size corresponding to misclassification.}
\label{fig1}
\end{figure}

To illustrate the concept, we apply our method to the simple example presented in Figure \ref{fig1}. For inputs $x \in [0,20]$, we set the true labelling function, $\Lambda$, to be the following:
\begin{equation}
\Lambda(x) = 
\begin{cases}
\hspace{0.2cm} l_{1} \hspace{0.5cm} \text{if} \hspace{0.2cm} x < 7 \\
\hspace{0.2cm} l_{2} \hspace{0.5cm} \text{if} \hspace{0.2cm} x \ge 7 .
\end{cases}
\end{equation}
The initial inputs are $\textbf{x} = \{0,1,3,5,6,8,11,12,15,17,19,20\}$, where inputs $ \textbf{x} = \{0,1,3,5,6\}$ are known to be in $R_{1}$ and are given label $l_{1}$, whilst the remaining points, $ \textbf{x} = \{8, 11, 12, 15, 17, 19, 20\}$, are known to be in $R_{2}$ and so are given label $l_{2}$. We have chosen a linear prior mean function for this example, and have applied a linear transformation to the inputs to help with the estimation of $\eta$. See Section \ref{prior} for a discussion on these prior choices.

The latent Gaussian process is estimated following the method outlined in Section \ref{meth}. Once we have a prediction for the labelling function, $\Lambda$, the boundary is estimated to be $7.4$, shown by the solid red line in Figure \ref{fig1}. This is a reasonable result since we set the boundary to be $x=7$, and we have not given the process any knowledge of where it actually lies between the two values. If the boundary actually was at $x=6$, the initial information we used to estimate the Gaussian process would not differ. This high level of uncertainty in our results is shown by the credible intervals for the estimate being large and are roughly equal to the bounds of $R_{1}$ and $R_{2}$ that we set the example up with.

\section{Misclassification} \label{miss}

The method of model validation we use is based on a leave-one-out cross-validation. In uncertainty quantification this usually involves leaving each training point out in turn, fitting a Gaussian process to the remaining points, and then using this to predict the point that was left out \citep{Seeger2004}. Given that our method models the class labelling function, $\Lambda$, we adapt this slightly to calculate a misclassification rate to see which of the initial inputs are more likely to be influential to our classification prediction. A leave-one-out cross-validation is performed on samples taken from $P(\eta(\textbf{x}) | \beta, \sigma^{2}, \delta, \Lambda(\textbf{x}))$ to predict the class label of each point left out in turn. From these samples, we calculate the proportion of times each point is misclassified. Points that have a large misclassification rate give an indication that the surrounding areas are likely to have a high uncertainty when making classifications. We expect points close to the boundary between $R_{1}$ and $R_{2}$ to have a high misclassification rate, and points where there are many neighbouring points (where we have high levels of information) to have a low misclassification rate. 

The output applied to the example is Section \ref{1de} in shown in Figure \ref{fig1}, where the size of the data points corresponds to the rate of misclassification. As expected, the rate is largest for the two points either side of the boundary. In a 1d example such as this, these points are the most critical since they restrict the boundary to the precise region of input space. It is also interesting to note here that the remaining points have a misclassification rate of almost (but not quite) zero. Taking a more in-depth look, we find that very occasionally the latent process crosses the axis. This is caused by the Gaussian process having a short correlation length parameter, leading to the latent process having the chance to bend quickly over the $\eta=0$ threshold between known points in the same region. This brings attention to the fact that prior choices can have a large effect on a method such as this with minimal initial information. We discuss this further in the following section.

\section{Prior choice methodology} \label{prior}

Based on the example in Section \ref{1de}, it is clearly important to place suitable priors on the model parameters and the prior mean function. One interesting aspect of Gaussian processes is their behaviour in the far edges of the input space. As Gaussian processes get far away from any data, they revert to the prior mean, $m$. This would be a problem for, say, a constant prior mean. If a constant mean function is placed on the Gaussian process, then we could start to observe the overall latent process tending towards the horizontal prior mean in the edges of our input space. Due to our model layout of $\eta$ being negative or positive according to the labelling, a constant prior mean may be estimated to be close to $m=0$, and we find that it is very easy for the process to switch signs, forcing a misclassification.

In many cases we will have extra information in our initial data that will help us to choose a more appropriate prior mean function. For example, we might use a prior mean function based on whether both regions are simply connected, or on the number of times the latent process is expected to change signs over the whole input space (for example by using a polynomial of that degree). Hence, any expert knowledge from the system modeller is very useful, particularly if they know how many distinct regions are expected. In a situation such as that in Figure \ref{fig1}, we have extra knowledge that there are only two output regions, and so it should be the case for the latent variable to only cross the $x$-axis once in the input space.  We used a linear mean function which forces the latent Gaussian process away from the $x=0$ axis in the edges of the input space. If, instead, one region was split either side of the other region, then it would be sensible to place a quadratic prior on the mean function, ensuring the Gaussian process would not return to the $x=0$ axis in the extreme values of the input space of the outer region. 

\begin{figure}[ht]
\centering
\includegraphics[scale=0.285]{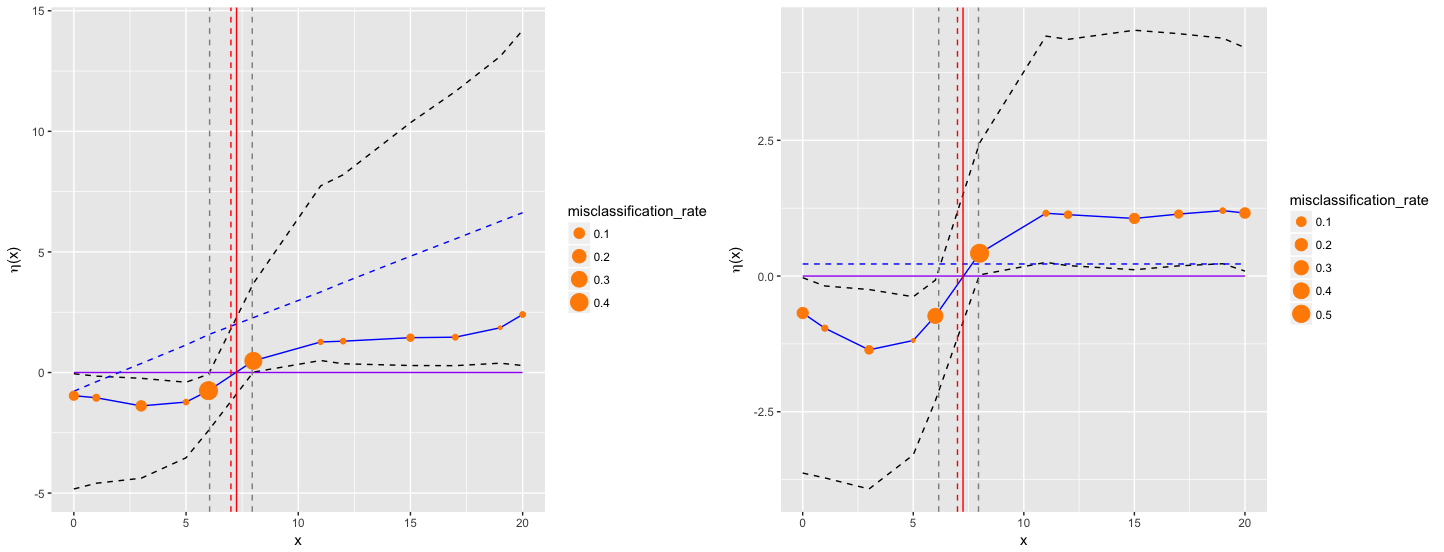}
\caption{Same example as in Figure \ref{fig1} but with some prior changes. The left plot is where the data are not transformed and the prior mean (red) crosses close to the origin (0,0). The right plot shows the effect of choosing a constant prior mean function.}
\label{fig2}
\end{figure}

The right plot in Figure \ref{fig2} shows the effects of choosing a constant mean function for the 1d example. Although the boundary estimate between $R_{1}$ and $R_{2}$ is almost the same as in the linear version in Figure \ref{fig1}, the latent process is significantly different. We can clearly see that the Gaussian process is returning to the prior mean (dashed blue) at the edges of the plot. The misclassification rates are also much larger here, confirming that we are much more likely to see misclassifications when using a constant mean function.

In two dimensions, if we had the area contained within a circle as one region, and the outer remaining space as the other region, then we could choose to fit a quadratic mean function. If instead we had two distinct circles as one of the regions and the outer remaining space was the other region, then we could consider fitting a quartic mean function. 

Although these would appear to be sensible choices, polynomials of a higher order come with a larger number of estimated parameters. Therefore, we should consider whether the classification in the edges of our input space away from the data is useful or not. It may be far more computationally expensive to calculate a high number of parameters and introduce more uncertainty than to assess whether the classification is accurate or not at the edges. As a rule of thumb, we would suggest choosing either a  linear or quadratic mean function if these are sensible choices. However, if it is decided that a higher polynomial would better suit the problem, but estimating the latent Gaussian process becomes far too complicated, then we suggest using a constant mean. When using a constant mean prior, we stress that it is important to be careful about any potential misclassifications, especially in the edges of the input region where a change in sign can be likely. If it is such that the number of data points is large, then we leave this up to the reader's decision on whether fitting a higher order polynomial is worthwhile or not.

If the choice is made to use either a linear or quadratic prior mean function, then there is more prior information that we can use to help with the computation. The estimated boundary between $R_{1}$ and $R_{2}$ becomes equivalent to the corresponding $\textbf{x} \in D$ such that it is the solution to $\eta(\textbf{x}) = 0$. Therefore, the latent process must cross the $x$-axis at approximately the boundary between $R_{1}$ and $R_{2}$, and we can encorporate this into the prior knowledge of our model. This will help approximate the parameter that controls where the latent process crosses the vertical axis more efficiently. 

As described in Section \ref{1de}, a transformation was applied to the input points so that the boundary between regions in the $x$-axis was approximately at zero in the vertical axis. With this transformation, a tight prior could be placed over the axis intercept parameter ensuring the latent process crosses the axis at zero. If we contrast the plot in Figure \ref{fig2} (left) compared with that in Figure \ref{fig1}, we notice a significant difference in the resulting latent process. The prior means for each plot are shown with dashed blue lines. Figure \ref{fig1} uses the transformed data and is shown to have an expected mean Gaussian process follow its prior direction. Figure \ref{fig2} (left) does not include the transformation and is shown to differ by the posterior estimate in region 1 leveling out as it approaches zero. This is clearly not appropriate since, in this simple two region example with no information of the system input, we would expect both sides of the latent process to match. This demnstrates that the transformation in the data can greatly improve the estimate in the latent process and any predictions that would follow.

When considering prior knowledge, it is also important to chose a good estimate for the correlation length parameter, $\delta$. The correlation length parameter determines how much the Gaussian process is allowed to bend between each of the initial data points \citep{Seeger2004}. If we consider the 1d example in Figure \ref{fig1}, we know that there is only one boundary where the latent Gaussian process is not expected to change sign between data points (apart from the boundary between regions). If the correlation lengths are allowed to become too small, then there is a chance that the Gaussian process would be able to curve round quickly and fall briefly in the wrong sign, causing a misclassification of regions in some input areas. To ensure this does not happen, inverse gamma priors are placed on the $\delta$'s so that they are forced away from zero and being too small. The mean of the prior distribution is kept away from being zero, whilst the scale is kept large to increase the spread. An inverse gamma prior is also placed on the variance, $\sigma^{2}$.

\section{Example in 2 dimensions} \label{2de}

\begin{figure}[ht]
\centering
\includegraphics[scale=0.85]{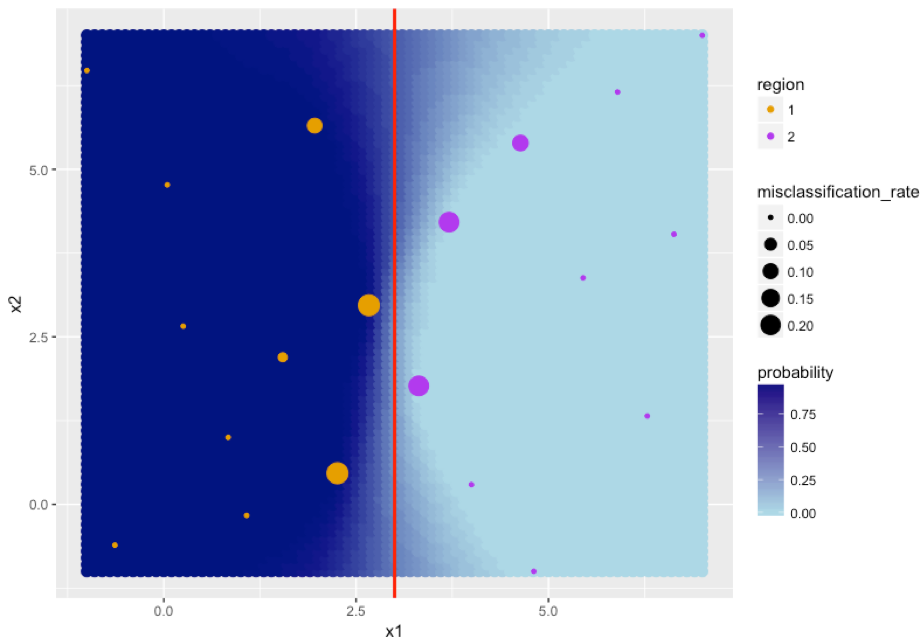}
\caption{2 dimensional example where the two region are split by an $x_{1}=3$ plane (red). The dark blue region corresponds to a high probability of be classified into $R_{1}$, whilst light blue corresponds to high probability of being classified into $R_{2}$. A misclassification rate is also shown based on point size.}
\label{fig3}
\end{figure}

Our method is now extended to include a simple 2 dimensional version of the 1d example from Section \ref{1de}. The output is shown in Figure \ref{fig3}, where 20 input points, $(x_{1},x_{2})$, are generated using a Latin hypercube \citep{Welch1992} over the region $[-1,7]^{2}$. We have specified the boundary between $R_{1}$ and $R_{2}$ in this example to be the line $x_{1}=3$ (shown in red) so that the true labelling function becomes:
\begin{equation}
\Lambda(x_{1},x_{2}) = 
\begin{cases}
\hspace{0.2cm} l_{1} \hspace{0.5cm} \text{if} \hspace{0.2cm} x_{1} < 3 \\
\hspace{0.2cm} l_{2} \hspace{0.5cm} \text{if} \hspace{0.2cm} x_{1} \ge 3 .
\end{cases}
\end{equation}

\begin{figure}[ht]
\centering
\includegraphics[scale=0.3]{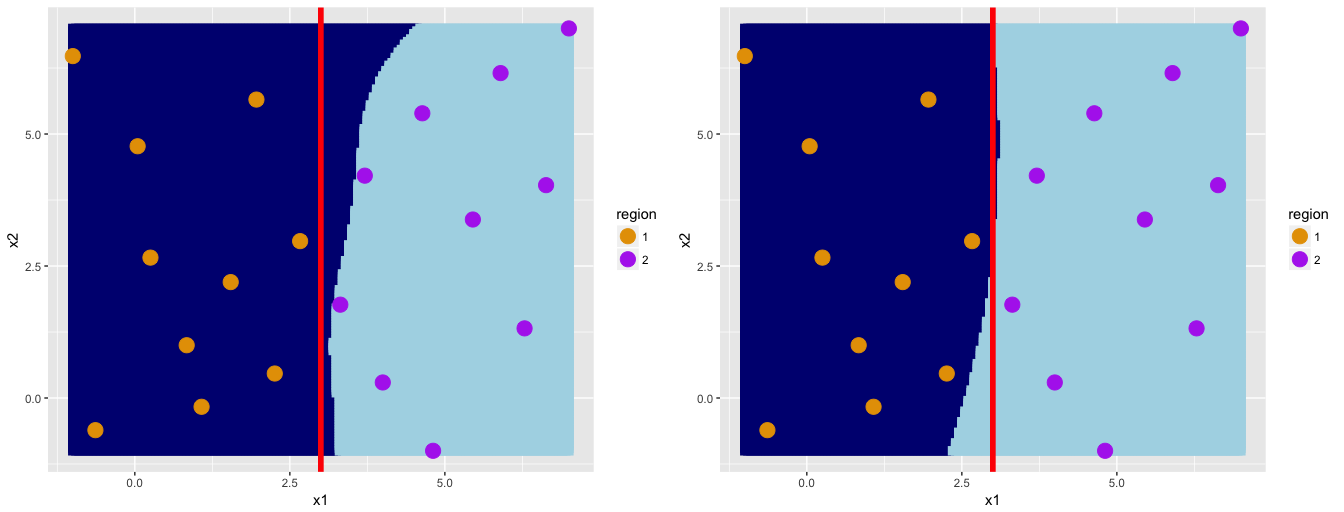}
\caption{Two different draws of the 2 dimensional example where the two region are split by an $x_{1}=3$ plane (red). The dark blue region corresponds to being classified into $R_{1}$, whilst light blue corresponds to being classified into $R_{2}$.}
\label{draws}
\end{figure}

From the plot, the yellow points are those data points in $R_{1}$ with label $l_{1}$ (input space $x_{1}<3$) and the purple points are those in $R_{2}$ with label $l_{2}$ (input space $x_{1}\ge 3$). The latent Gaussian process has been applied to a grid of points over the input space to show the estimated classification labellings. Two possible draws from the latent Gaussian process are shown in Figure \ref{draws}. 

To show uncertainty within the 2d example, Figure \ref{fig3} shows the probability of input points being classified into $R_{1}$ compared with $R_{2}$. The dark blue regions represent high probability of being classified into $R_{1}$ and the light blue represents high probability of being classified into $R_{2}$. A misclassification rate is calculated for each point as described in Section \ref{miss}, and is shown in Figure \ref{fig3}. As expected, the points near the boundary have a larger rate of misclassification and the uncertainty increases. 

We chose to fit the Gaussian process with a linear prior mean because we know that the boundary is vertical in the x-axis. Even if we didn't know this apriori, we can see from the initial data that a linear mean might be a reasonable choice since there only seems to be one change in region over the space filling design.

\section{Comparison with existing methods} \label{comp}

Using the example in the above section, we now illustrate the strengths of our method by comparing it with existing methods. One of the most widely used methods for this type of classification is logistic regression \citep{Diggle1998, Kleinbaum1994, Hilbe2009}. Similar methods in uncertainty quantification literature can also be seen by \cite{Chang2015} and \cite{Salter2018}. 

\begin{figure}[ht]
\centering
\includegraphics[scale=0.3]{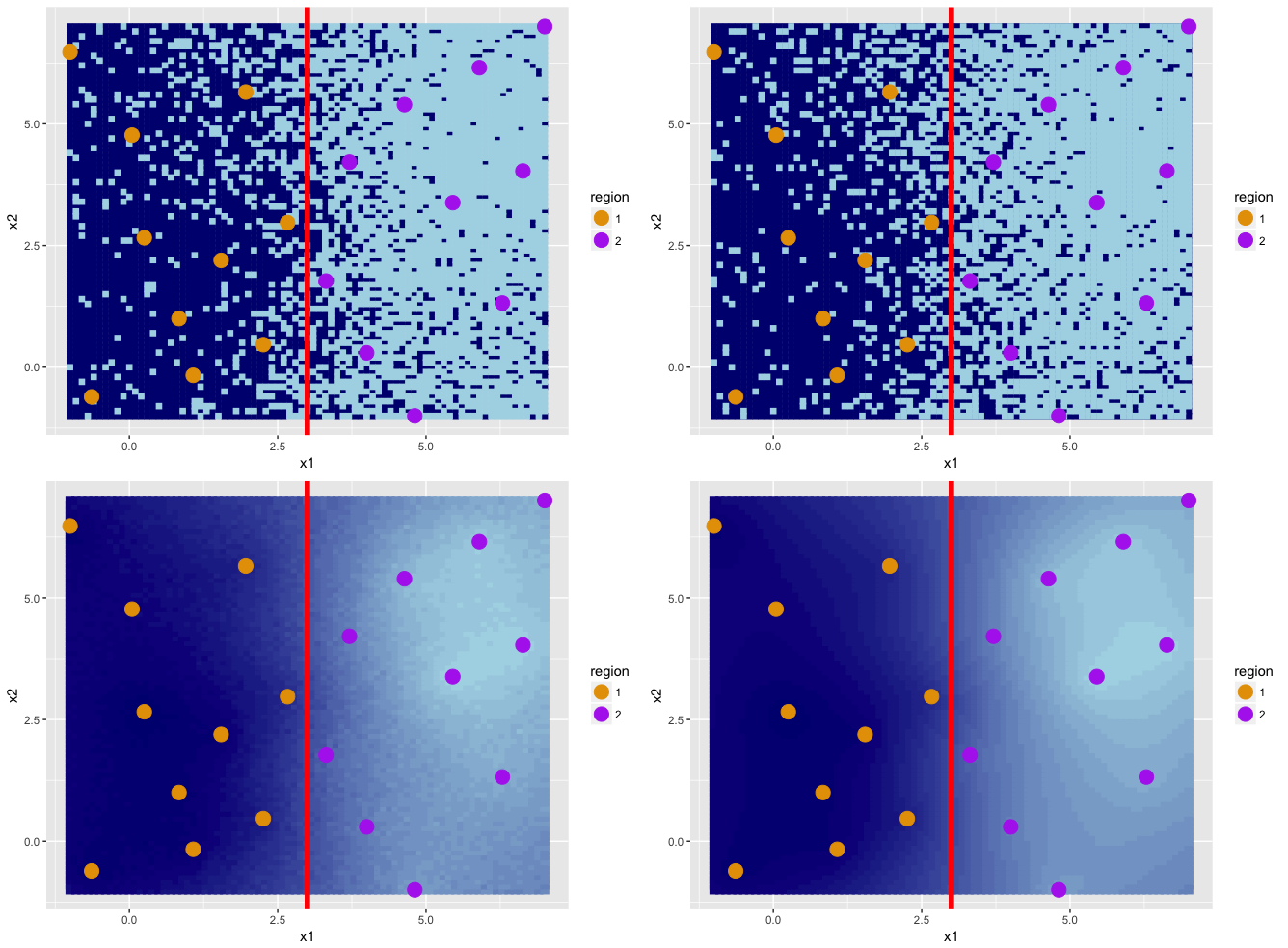}
\caption{2 dimensional example from section \ref{2de} modelled using logistic regression. Top row: Bernoulli samples of region classifcations using logistic regression. Bottom left: average of 1000 Bernoulli samples. Bottom right: underlying probability function of being classified into $R1$ or $R2$.}
\label{logreg}
\end{figure}

The outputs of using logistic regression for this 2d example are shown in Figure \ref{logreg}, where the logistic model is stated to be the following:
\begin{equation}
\
\Lambda(\textbf{x}) \sim \text{Bernoulli}(\eta(\textbf{x})), \hspace{2cm}
\text{logit} \left( \eta (\textbf{x}) \right) = \beta_{0} + \beta_{1} x_{1} + \beta_{2} x_{2},
\end{equation}
where $\Lambda(.)$ is the labelling function and $\eta(.)$ is the latent Gaussian process. The bottom right plot shows the underlying probability of being classified into $R_{1}$. This is a smooth function that shows high probability of being sorted into either of the regions where expected. Two samples are shown in the top two plots in Figure \ref{logreg}, which have been computed using the geoRglm package in R. This exposes the main flaws of using logistic regression for this framework. We can clearly see that because the random Bernoulli sample does not take into account the distance correlation between points, the sampled $\Lambda(\textbf{x})$ field is not smooth. In practice, $E[\Lambda(\textbf{x})]$ might be used as a classifier, but with our method, our samples all give a coherent full classification. There is no distinct boundary between $R_{1}$ and $R_{2}$. Comparing to samples drawn using our method in Figure \ref{draws}, we are able to produce a clean cut boundary in every sample. Also, \cite{Chang2015} and \cite{Salter2018} fit a Gaussian process to $\eta(\textbf{x})$, but they use a Bernoulli likelihood, which is not used here.

The main problem is the lack of correlation when sampling from the Bernoulli distribution, where our method is able to classify jointly over the input space. The underlying probabilty function retains the correlation structure, but when we sample marginally over all points, all of this is lost. We could run a smoother over these samples, but this can get very complicated, and we still would not be able to define the exact boundary between regions. Alternatively, we could take a threshold on the probability function, but it is not clear what value we would choose for this threshold. To make a fairer comparison to our method, the bottom left plot in Figure \ref{logreg} averages over 1000 Bernoulli samples. This now has the smoothness of the probability function, and is similar to our plot in Figure \ref{fig2}, but still does not provide an estimate of the boundary. Due to this boundary estimate, we are able to classify a set of points at the same time rather than only individual points. It is also important to note that if averages are all that is required then logistic regression does a reasonable job. There are however cases when we require samples, which is where our method clearly outperforms.

\begin{figure}[ht]
\centering
\includegraphics[scale=0.3]{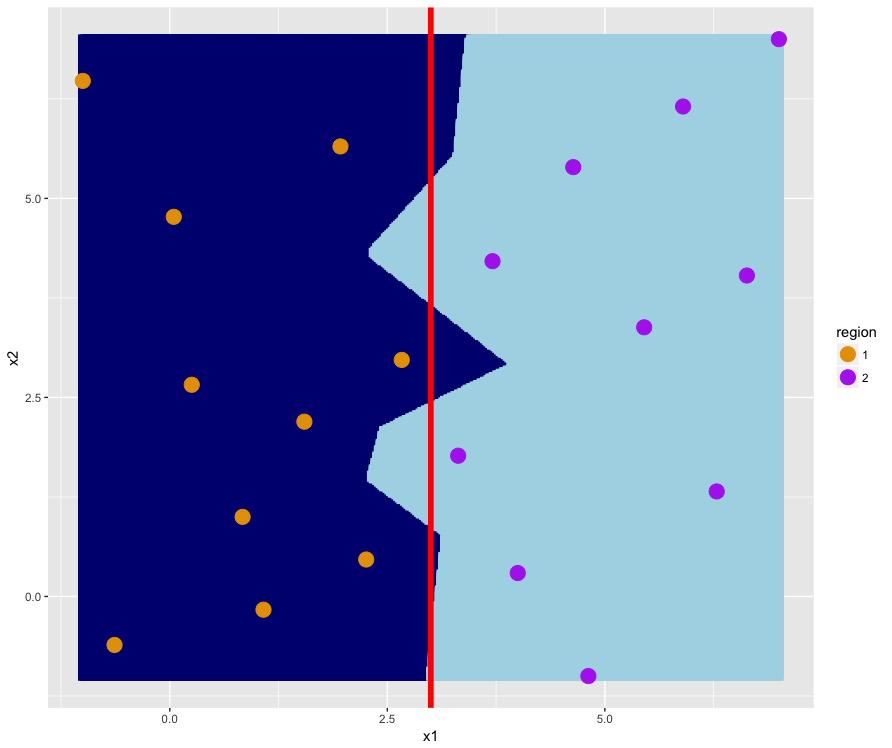}
\caption{2 dimensional example from section \ref{2de} where classifications have been made using Voronoi tessellations.}
\label{naive}
\end{figure}

Figure \ref{naive} shows a naive approach to this problem by splitting up $R_{1}$ and $R_{2}$ using Voronoi tessellation \citep{Gallier2008, Kim2005}. We can clearly see that this outperforms the classifications made using logistic regression, but again has several flaws. One of the main problems is that it is only able to classify along straight lines, which sacrifices a certain amount of precision compared to our method. Another point to make is that with Voronoi tessellation, we just take the mid-point between the closest points in $R_{1}$ and the closest points in $R_{2}$. Our method is able to learn more from all initial points. This is shown in Figure \ref{fig2} where the upper section of our boundary estimate is shown to curve far more into $R_{2}$ than that of the lower section. Similarly, we can not make any uncertainty statements with this method. 

\section{Another example in 2 dimensions} \label{Sant}

\begin{figure}[ht]
\centering
\includegraphics[scale=0.4]{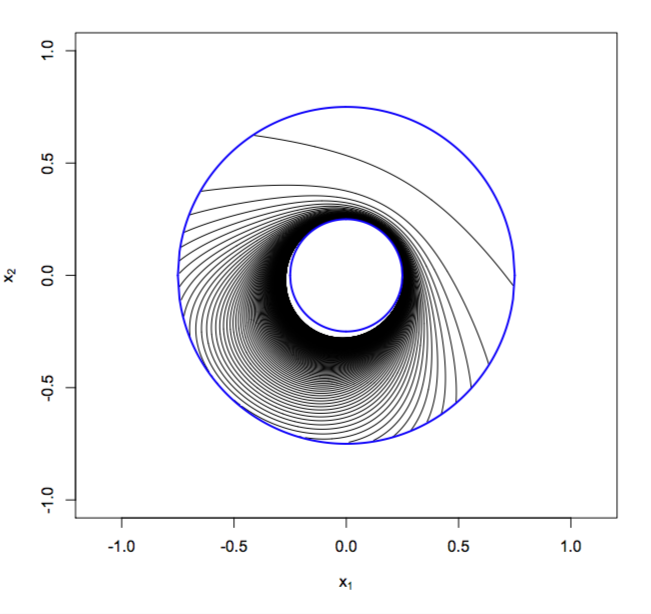}
\caption{2 dimensional example with two regions. $R_{1}$ lies within the two circles and $R_{2}$ is the remaining input space. The contours show the function, $f$, for various values of $x_{1}$ and $x_{2}$.}
\label{fig8}
\end{figure}

We apply our method to an example provided by T. Santner, with test function:
\begin{equation}
f(x) = 
     \begin{cases}
       \infty &\quad \text{if} \hspace{0.2cm} x_{1}^{2} + x_{2}^{2} \le c_{1}^{2} \\
       \frac{\exp{-(a'x + x'Qx)}}{(x_{1}^{2} + x_{2}^{2} - c_{1}^{2})} &\quad \text{if} \hspace{0.2cm}  c_{1}^{2} \le x_{1}^{2} + x_{2}^{2} \le c_{2}^{2} \\
       - \infty &\quad \text{if} \hspace{0.2cm} x_{1}^{2} + x_{2}^{2} \ge c_{2}^{2} ,\\
     \end{cases}
\end{equation}
where,
\begin{equation}
a = [3,5] \hspace{0.5cm} 
Q = 
\left(\begin{array}{cc} 
2 & 1.5 \\
1.5 & 4  \\
\end{array}\right)
\hspace{0.5cm} c_{1}^{2} = 0.25^{2}, \hspace{0.3cm} c_{2}^{2} = 0.75^{2} \hspace{0.1cm}.
\end{equation}
This function is plotted in Figure \ref{fig8}.

The space between the two circles is $R_{1}$ and the remainder is $R_{2}$, both over the input space $[-1.25,1.25]^{2}$. The output function to the model, $f$, is only valid for $R_{1}$, so the true labelling function, $\Lambda$, becomes:
\begin{equation}
\Lambda(x_{1},x_{2}) = 
\begin{cases}
\hspace{0.2cm} l_{1} \hspace{0.5cm} \text{ if } \hspace{0.2cm} 0.25^{2} \le x_{1}^{2}+ x_{2}^{2} \le 0.75^{2} \\
\hspace{0.2cm} l_{2} \hspace{0.5cm} \text{ if } \hspace{0.2cm} x_{1}^{2}+ x_{2}^{2} < 0.25^{2} \text{ OR } x_{1}^{2}+ x_{2}^{2} > 0.75^{2} .
\end{cases}
\end{equation}

\begin{figure}[ht]
\centering
\includegraphics[scale=0.8]{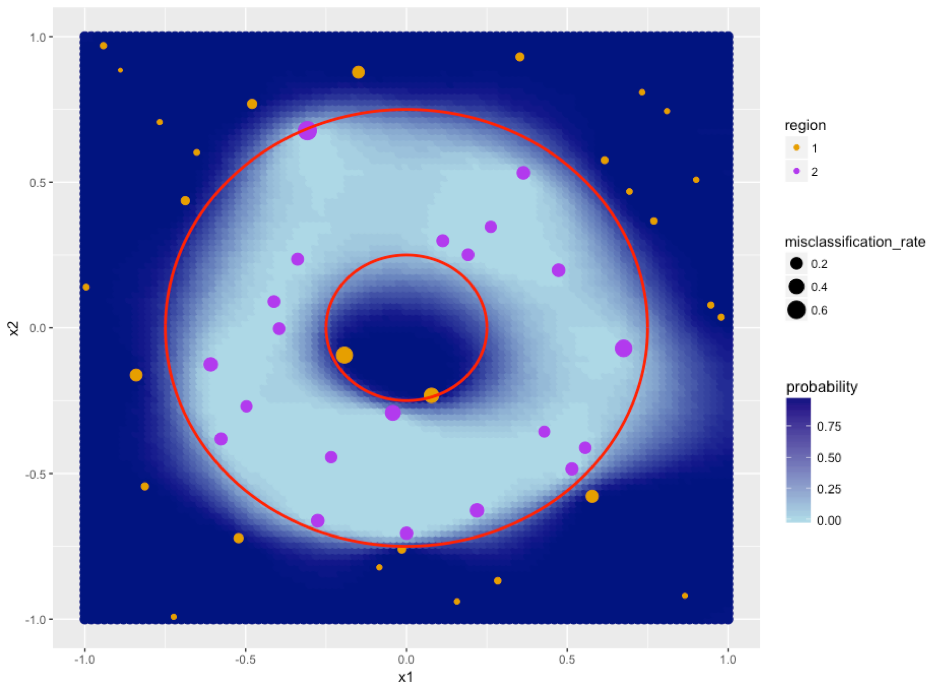}
\caption{Estimated regions for the 2d example shown in figure \ref{fig8}. Initial data points are displayed (orange - region 1 and purple - region 2), with the actual region boundaries shown in red. Uncertainty on the estimate is included where light blue areas correspond to high probability of being classified into $R1$ and dark blue areas correspond to high probability of being classified into $R2$. A misclassification rate is also shown.}
\label{fig7}
\end{figure}

\begin{figure}[ht]
\centering
\includegraphics[scale=0.22]{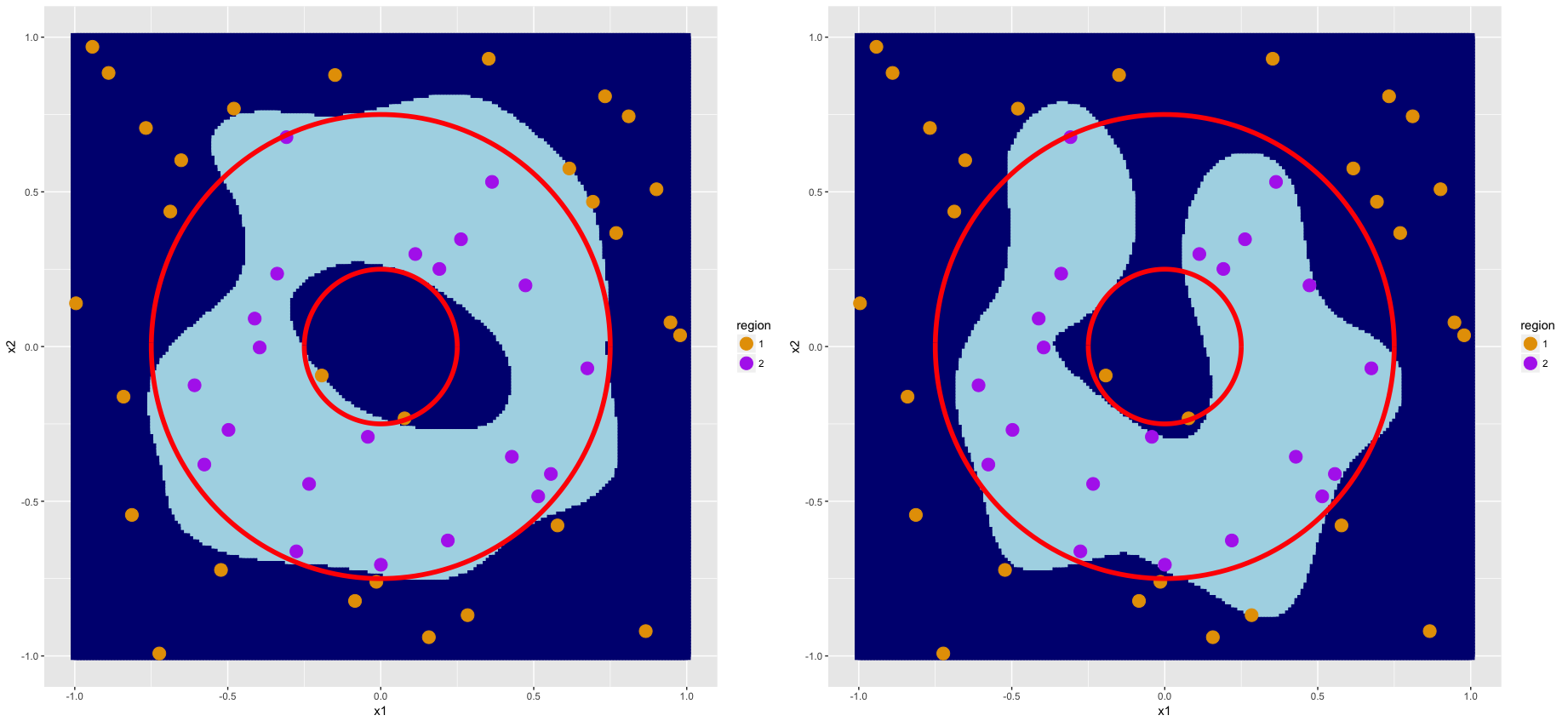}
\caption{Two different draws from the 2 dimensional example with two regions. $R_{1}$ lies within the two circles and $R_{2}$ is the remaining input space. The dark blue and light blue regions correspond to areas being classified into $R_{1}$ and $R_{2}$ respectively.}
\label{TSsamps}
\end{figure}

We have used a 2d maximin Latin hypercube to select 50 data points, $(x_{1},x_{2}) \in D$, where they are given class labels $l_{1}$ if $(x_{1},x_{2}) \in R_{1}$ and $l_{2}$ if $(x_{1},x_{2}) \in R_{2}$. These are shown by the purple and orange points in Figure \ref{fig7} along with the hard boundary (red). The labelling classification after applying our method is also shown in the plot with uncertainty as the background colour. As in the previous example, the light blue areas represent a high probability of being labelled $l_{1}$ and the dark blue areas show high probability of being labelled $l_{2}$. The largest areas of uncertainty correspond to the areas where our classification method performed the poorest. Figure \ref{TSsamps} shows two draws from the latent GP. The plot on the left is fairly accurate to the truth, but it is particularly interesting to note that the doughnut shape in the right plot is no longer fully connected. This is likely to be due to a lack of information in that area of input space. 

Overall, our method is estimating the regions well with only a few larger deviations in the upper left and right sections of the doughnut region. This is likely to be caused by a lack of information in these areas. Due to the more complicated shape, we chose to fit a constant prior mean function. This has proven to be successful, since no areas have been misclassified in the far corners of the input space. Alternatively, a quartic polynomial could have been used for the prior mean function, but, as mentioned in Section \ref{prior}, we do not recommend using anything with greater complexity than a quadratic (unless there is a sufficient quantity of data). 

Two input points that produce interesting results are those at the bottom of the larger circle; they are classified in different regions but are very close together. In this area, the latent Gaussian process must change sign quickly but has been able to without any complications.

A misclassification rate is also included, where the points are more likely to missclassify in $R_{1}$ (between rings). This is likely to be due to a higher proportion of points being in $R_{2}$ and so the majority of the latent process is negative, making it more likely for areas to be classified into $R_{2}$. This is supported by the constant mean function estimated to be $-2.25$. 

We can also attribute the larger misclassification rates (compared to those in Figure \ref{fig3}) to the use of a constant mean function; it becomes a lot more uncertain without the directional force of a higher order polynomial. Although it appears that the corners misclassify very infrequently, when we observe the underlying latent Gaussian process (not shown), $\eta$, we can see that it is infact starting to curve up in the corners towards the constant mean value.

\section{Application} \label{app}

Our motivating example has been supplied by \cite{Voliotis2018}, where the subject is the reproductive system in mammals, particularly how it is controlled by connections between the brain, the pituity gland, and the gonads. There are particular neurones in the brain that secrete a specific hormone known as the gonadotrophin-releasing hormone (GnRH). These are vital in regulating gametogenesis and ovulation. Signals are made by the pituitary gland which then simulate the gonads for this cycle to start. One of the regulators of the GnRH neurone is neuropeptide kisspeptin, of which two are located within areas of the hypothalamus (the arcuate nucleus (ARC) and the proptical area). Other research suggests that one of these areas (ARC) is the location of the GnRH pulse regulator of which the core are neurones (ARC kisspeptin or KNDy) that secretes two neuropeptides: neurokinin B (NKB) and dynorphin (Dyn). The object of the model presented is to understand the role of NKB and the firing rate of these neuropeptides on the regulation of GnRH, and subsequentially in controlling reproduction. To do this, the model identifies the population of the KNDy neurones where the GnRH pulse regulator is said to be found. The model consists of a set of coupled ordinary differential equations (ODEs) to describe the dynamics of $m$ synaptically connected KNDy neurones. There are several fixed parameters, including the concentration of Dyn, rates at which Dyn and NKB are lost and those that describe the characteristic timescale for Dyn and NKB. The variables are the concentration of NKB secreted at the synaptic ends and the firing rate, measured in spikes/min. Using the population of KNDy neurones is shown to be critical for GnRH pulsatile dynamics and that this can stimulate GnRH secretion. Analysing the output of this model shows that the population can behave as a bistable switch so that the firing rate is either high or low. Hence, this causes us to have a system with two distinct solutions, and is an example of the type of system that we wish to model. This bistable system is coupled with a negative feedback leading to sustained oscillations that drive the secretion of GnRH hormones that are involved in reproduction. Being able to model the system and locate the areas of low and high firing rates means that, not only can we aide predictions on the repreduction rate, but we can also have a better understanding of the specific input parameters that are associated with high rates of reproduction.

\begin{figure}[ht]
\centering
\includegraphics[scale=0.85]{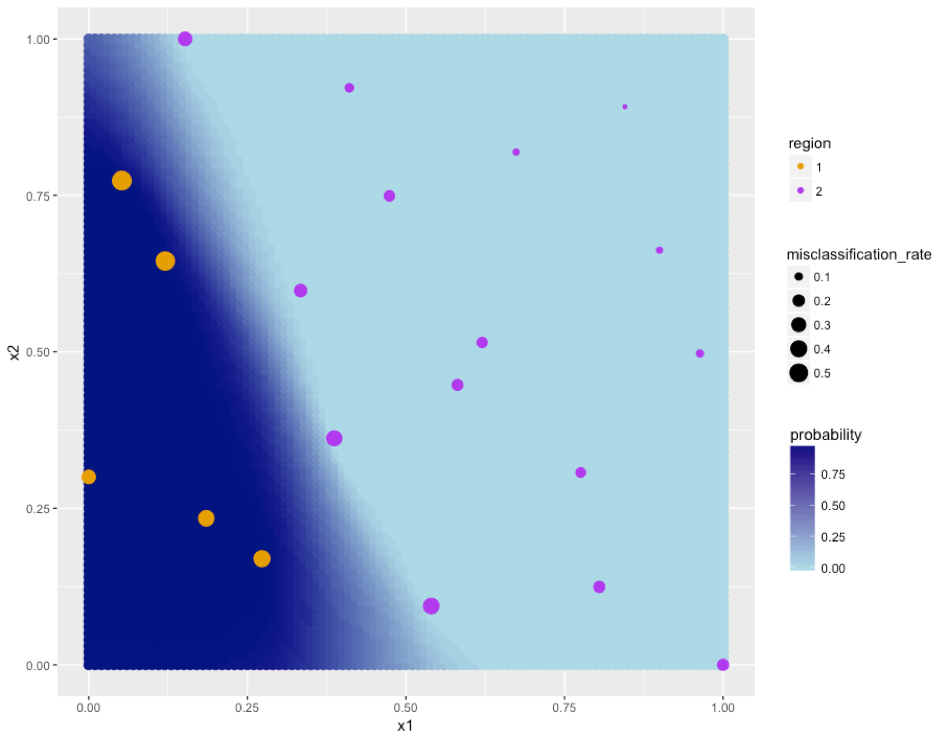}
\caption{2 dimensional example looking at the effects of hormone release on mammal reproduction, where the system has two regions of high and low rates of hormone release. Initial points are displayed (orange - $R_{1}$ and purple - $R_{2}$), with predicted region classification and uncertainty. Dark blue areas correspond to high probability of being classified into $R_{1}$ and light blue areas correspond to high probability of being classified into $R_{2}$. A misclassification rate is also shown.}
\label{fig5}
\end{figure}

\begin{figure}[ht]
\centering
\includegraphics[scale=0.21]{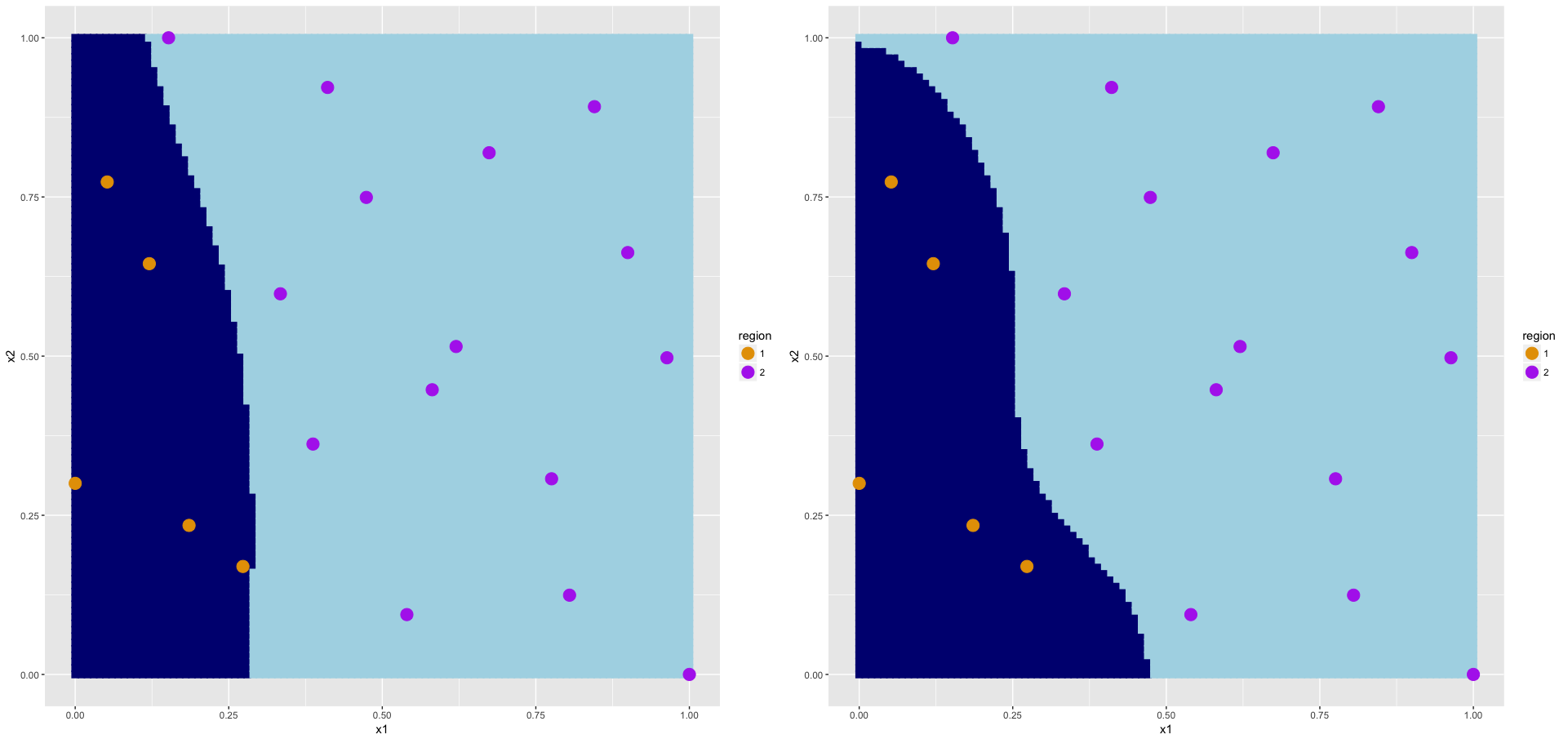}
\caption{Two different draws from the 2 dimensional application with two regions. Initial inputs are shown in yellow and purple with the dark blue and light blue regions corresponding to areas being classified into $R_{1}$ and $R_{2}$ respectively.}
\label{expsamps}
\end{figure}

The inputs are NKB concentration and firing rate, where we create a Latin hypercube over the input space of $[0.1,0.2] \times [10,200]$. The choice was made here to transform the data to $[0,1]^{2}$ for computational simplicity. The system is bimodal, so for 20 initial points where we know the region classification, we can apply our labelling function, $\Lambda$. We have 5 points labelled $l_{1}$ in $R_{1}$, and 15 as $l_{2}$ in $R_{2}$ (as seen in Figures \ref{fig5} and \ref{expsamps}). The true function, $\Lambda$ and resultant boundary are not known in this example.

One of the most important choices to be made in this example was the form of the prior mean on the latent Gaussian process. We chose a linear prior based on consultation with the expert of the system and examination of the initial points (yellow and purple shown in Figure \ref{fig5}). The output of the predicted region boundary is shown in Figure \ref{fig5}, as well as the uncertainty. In general, our solution classifies as expected in most areas, where the area between the regions is the most uncertain. We would therefore expect the true boundary between $R_{1}$ and $R_{2}$ to be almost a straight line, with potential to curve at either of the ends of the input space. This uncertainty is down to lack of information at the boundaries, but since our area of uncertainty is not too large, we have greater confidence in our estimation. Figure \ref{expsamps} shows two draws from the latent GP and confirms that there is more uncertainty in the lower half of the input space close to the boundary. They both capture a similar linear trend, with the plot on the right having more curvature. Misclassification is shown by the size of the points, and we see that is is easier to misclassify points near the boundary.  

\section{Discussion} \label{con}

We have developed a new method for classifying models or simulators where the output is labelled according to two regions. Our method includes correlation through a distance metric, and it can be applied to a broad range of applications where outputs from the model are not necessarily quantitive. A major disadvantage of most common methods of classification, such as logistic regression, is the assumption that class labels associated with input points are independently distributed Bernoulli random variables. This is something that causes concern since any correlation between nearby points is ignored. Neighbouring input points are more likely to result in the same output label, so it is vital that we include this information in our model, particularly if we are feeding samples into a more complex statistical model.

Keeping this in mind, we used aspects of classification from \cite{Nickisch2008} in the form of class labelling and incorporate Gaussian process emulation. To ensure that correlation between data points was included, a latent variable modelled as a Gaussian process is used to structure the two output solutions using our assigned class labelling. The latent Gaussian process is estimated using MCMC with distinct prior specifications. As a form of model validation, we have calculated a misclassification rate which is based on a leave-one-out cross-validation.

We feel that this method will be applicable to a wide range of applications across many disciplines including computer science, climate science and biology. Our main motivating example is based on assessing reproduction rates in mammals \citep{Voliotis2018}. We have successfully modelled this bimodel system, for which our model can be used for class prediction for other input points with estimates of uncertainty included. Comparisons have also been made with logistic regression and voronoi tessellation.

There are some obvious extensions to the work presented in this paper. One would be to now expand the method to cope with situations when there are more than two output labels. This would then increase the numer of applications where it is suitable. There is also room for research in areas of experimental design where we can improve the accuracy of our class classification and boundary estimation with limited initial data.

\section*{Acknowledgements}
The authors gratefully acknowledge T. Santner and M. Voliotis for providing the given examples. Louise Kimpton would also like to thank EPSRC for her studentship.

\section*{References}

\bibliography{REF2}

\end{document}